\documentclass[a4paper, 12pt, twocolumn]{article}

\usepackage[top=75pt,bottom=75pt,left=40pt,right=40pt]{geometry}

\usepackage[utf8x]{inputenc}
\usepackage{layout}
\usepackage{color}
\usepackage[dvipsnames]{xcolor}
\usepackage{lipsum}
\usepackage{graphicx}

\usepackage[small]{caption}
\usepackage{subcaption}
\usepackage{gensymb}
\usepackage{amssymb}
\usepackage{mathtools}
\usepackage{xspace}
\usepackage{upgreek}
\usepackage{enumerate}

\usepackage{balance}

\usepackage[switch]{lineno}

\title{\bf {\color{MidnightBlue} Measurement of the muon flux in the bunker of Monte Soratte with the CRC detector}}
\author{A.~Candela\textsuperscript{1},  
A.~Cocco\textsuperscript{1}, 
N.~D'Ambrosio\textsuperscript{1}, 
M.~De Deo\textsuperscript{1}, 
A.~De Iulis\textsuperscript{3}, \\
M.~D'Incecco\textsuperscript{1}, 
P.~Garcia Abia\textsuperscript{4}, 
C.~Gustavino\textsuperscript{2}\textsuperscript{*},
G.~Gustavino \textsuperscript{5}, \\ 
M.~Messina\textsuperscript{1}, 
G.~Paolucci\textsuperscript{3},  
S.~Parlati\textsuperscript{1} and 
N.~Rossi\textsuperscript{1} 
}
\date{\footnotesize \today}

\begin{document}

\twocolumn[
\begin{@twocolumnfalse}
\maketitle

\vspace{-1cm}
\begin{center}
\footnotesize \it \textsuperscript{1}Laboratori Nazionali del Gran Sasso, Assergi (AQ), Italy \\
\footnotesize \it \textsuperscript{2}Istituto Nazionale di Fisica Nucleare, Sezione di Roma, Italy; \\
\footnotesize \it\textsuperscript{3}Museo Percorso della Memoria, Sant'Oreste (RM), Italy; \\
\footnotesize \it\textsuperscript{4}CIEMAT, Centro de Investigaciones Energ\'eticas, 
\footnotesize \it Medioambientales y Tecnol\'ogicas, Madrid, Spain; \\
\footnotesize \it \textsuperscript{5}Homer L. Dodge Department of Physics and Astronomy, 
\footnotesize \it University of Oklahoma, Norman OK, USA. \\
\footnotesize \textsuperscript{*}email: \texttt{carlo.gustavino@roma1.infn.it}
\end{center}

\begin{abstract}
In the context of the \textsc{Ptolemy} project, the need for a site with a rather low cosmogenic induced background led us to measure the differential muon flux inside the bunker of Monte Soratte, located about 50~km north of Rome (Italy). The measurement was performed with the Cosmic Ray Cube (CRC), a portable tracking device. The simple operation of the Cosmic Ray Cube was crucial to finalise the measurements, as they were carried out during the COVID-19 lockdown and in a site devoid of scientific equipment.
The muon flux measured at the Soratte hypogeum is above two orders of magnitude lower than the flux observed on the surface, suggesting the possible use of the Mt. Soratte bunker for hosting astroparticle physics experiments requiring a low environmental background. 
\end{abstract}

\vspace{1cm}

\end{@twocolumnfalse}
]

\maketitle

\section*{Introduction}

The \textsc{Ptolemy} project aims at the detection of cosmic neutrino background (CNB)~\cite{Baracchini:2018wwj,Betti:2019ouf,Betti:2018bjv} produced in the early universe according to the Big Bang theory~\cite{BBC}.
Due to the low reaction rate expected, this experiment would operate in a site with low environmental radioactivity, suggesting its underground installation to reduce the cosmic-ray induced background.

A candidate site to host the \textsc{Ptolemy} experiment, as well as other astroparticle physics experiments~\cite{cygnus}, is the bunker of the Soratte mountain, at about 50 km north of Rome, Italy.
Its construction started in 1937 as an air-raid shelter. During the 1967-1972 period its structure was modified to realize an anti-atomic site. Presently, the bunker is composed of a network of tunnels and halls for a total length of about 4 km and it hosts the ``Path of memory'' historical museum~\cite{museo}.

The integrated flux of vertical muons with energy above 1~GeV/c at sea level is $70~\mathrm{m^{-2} s^{-1} sr^{-1}}$~\cite{pdg,pdg56,pdg57}. Recent measurements~\cite{pdg51,pdg58,pdg59} favor a 10-15\% lower normalization. The integral intensity of muons in the Soratte bunker is expected to be significantly lower.
Similar measurements are performed in shallow-underground laboratories~\cite{Felsenkeller}.
The muon flux measurement discussed in this paper is performed in the ``War Room'' of the museum with the Cosmic Ray Cube (CRC), a novel detector made mainly for educational purposes.

The paper is organized as follows: Section~\ref{sec:CRC} describes the CRC detector, the detector performance and the event reconstruction are discussed in section~\ref{sec:reco}, and section~\ref{sec:muon} presents the data analysis and the results.

\section{The CRC detector}
\label{sec:CRC}

The CRC detector is the evolution of the hodoscope exhibited in L'Aquila in 2009 during the Group of Eight (G8) summit meeting and afterwards installed in the Museum of Science (Teramo, Italy). That device consisted of 10 planes of Glass Resistive Plate Chamber (GRPC) equipped with orthogonal read-out strips. Its peculiar feature was the ability to visualize the cosmic-ray tracks with light-emitting diodes (LEDs) connected to the front-end electronics~\cite{compact}.
A more recent version of a visual hodoscope for cosmic rays is installed in the subway station ``Toledo'' (Naples, Italy) and other sites~\cite{adriano}.
The most relevant improvement of this device is the use of scintillating bars coupled to silicon photomultipliers (SiPM) instead of GRPC detectors, making the device more suitable for outreach purposes. The CRC is an simplified version of the previous hodoscopes. It is a portable device, easy to construct by assembling a few basic components, that requires only the standard electrical power and Internet connection for its operation.

\begin{figure}[ht]
   \centering 
   \includegraphics[width=1.0\columnwidth]{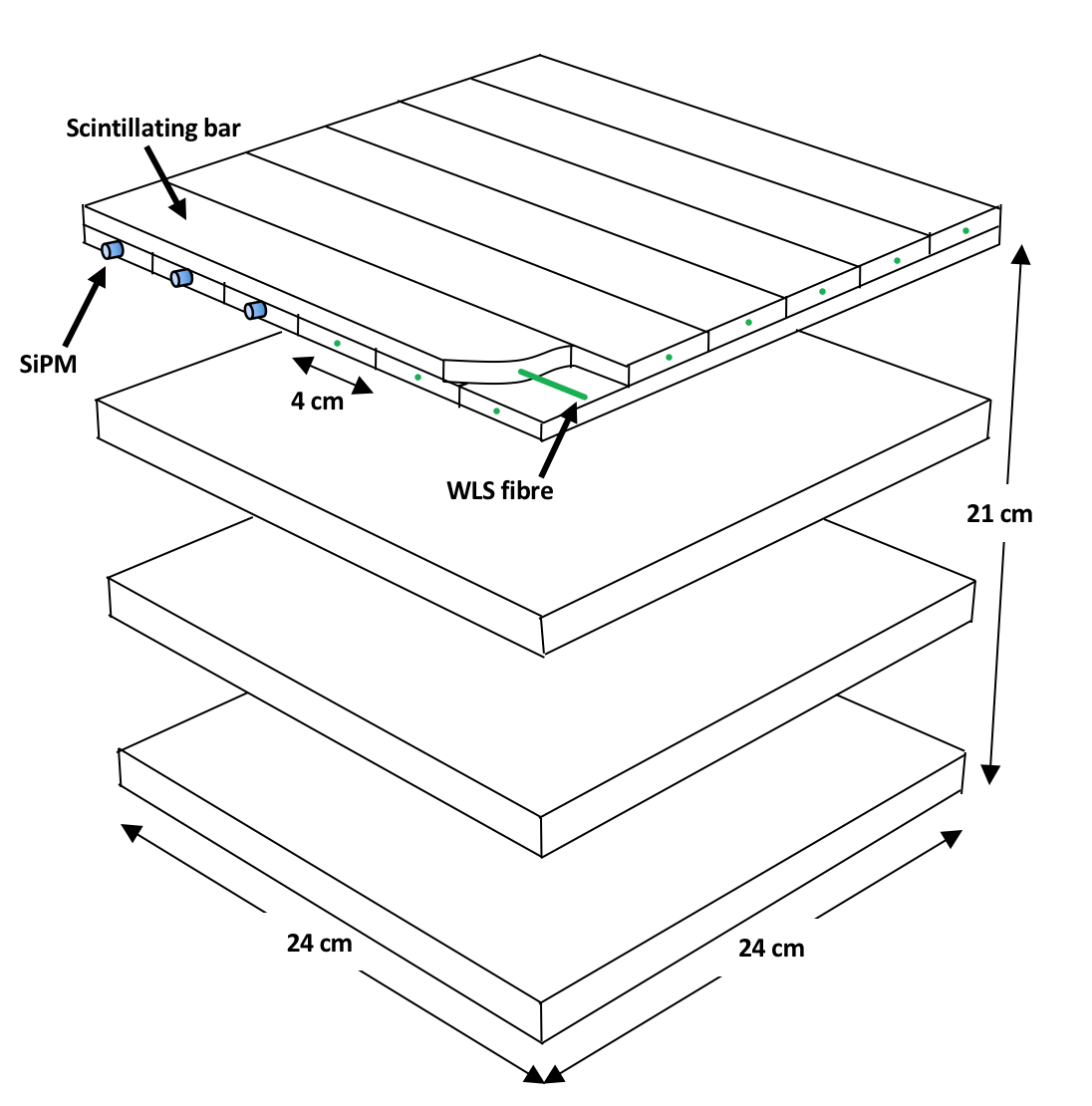}
   \caption{Schematic drawing of the CRC detector.}
   \label{fig:sketch}
\end{figure}

\begin{figure}[ht]
   \centering
   \includegraphics[width=1.0\columnwidth]{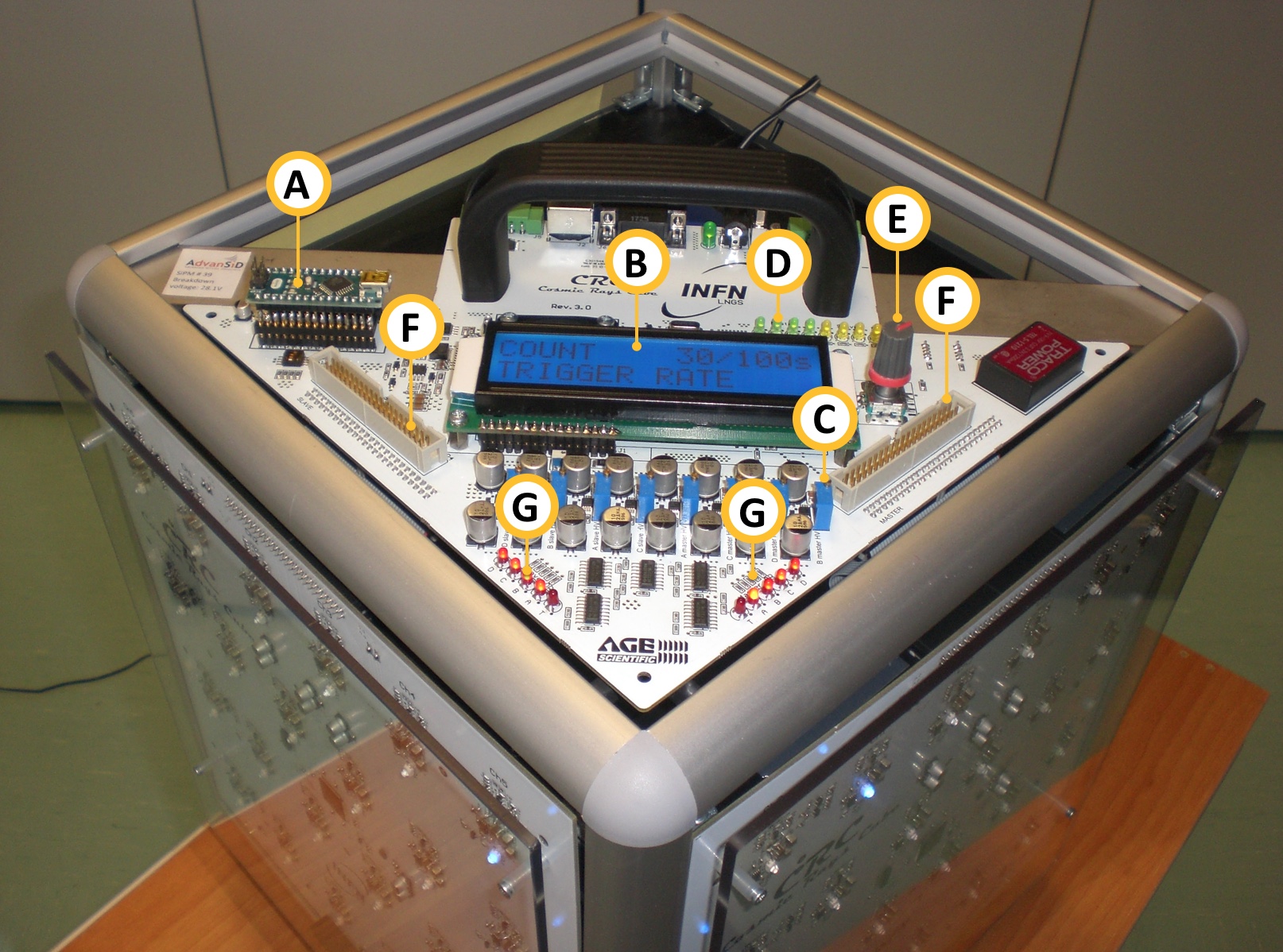}
   \caption{CRC controller board. Arduino based trigger acoustic alert (A), display of set function (B), SiPM Bias voltage adjustment  (C), trigger configuration LEDs (D), encoder to select menu function and value of selected parameters (E), voltage and signal probes (F),x counting rate LEDs (G).}
   \label{fig:scheda}
\end{figure}

The CRC is based on the use of STYRON 663 scintillating bars arranged as shown in figure~\ref{fig:sketch}. 
The bars are 24~cm long with a rectangular section of $4 \times 1~\mathrm{cm}^2$. They are coated with titanium dioxide to maximize the optical signal and prevent noise from external light~\cite{scint}.
Each bar is coupled with a Kuraray Y-11(200) wavelength shifting (WLS) fibre that collects and transmits the light to a Silicon PhotoMuptiplier (SiPM) (AdvanSiD ASD-RGB1C-M) that converts the photons into an electrical signal. 
The real-time visualization of cosmic rays is provided by LEDs placed at the end of each bar. In total, the CRC is composed of four $x$-planes and four $y$-planes, each one consisting of 6 bars, for a total of 48 electronic channels (figure~\ref{fig:sketch}).

The detector is equipped with three printed circuit boards (PCB), two of which are implemented with an Altera Max V Complex Programmable Logic Device (CPLD) for signal processing. The $3^\mathrm{rd}$ card unit (CONTROLLER) allows the setting of the signal thresholds, the SiPM bias voltages and the trigger configuration. In addition, it manages the data and provides the power supply to operate the CRC. The onboard display and the encoder make it possible to define different functions such as the trigger configuration and the monitoring of channels rates (figure~\ref{fig:scheda}). A microchip controller is used to manage the digital pattern of events, that can be registered in a file together with their trigger time.

\begin{figure}[ht]
   \centering
   \includegraphics[width=1.0\columnwidth]{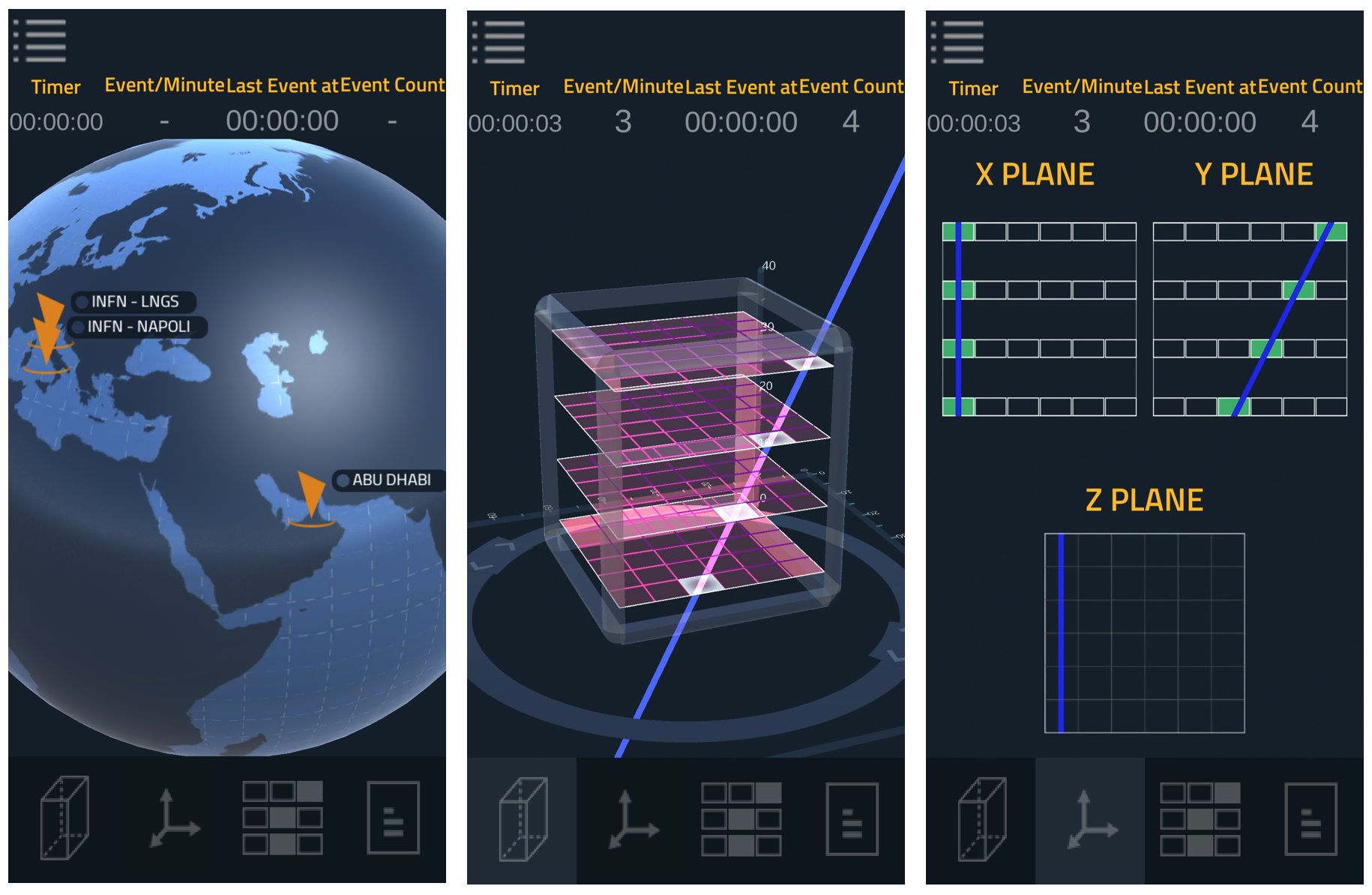}
   \caption{Screenshots of the ``Cosmic Ray Live'' application for smartphones. From left to right: worldwide locations of CRC sites, 3D track reconstruction, projections of a muon track.}
   \label{fig:app}
\end{figure}

The data files and several real-time functions, such as the event display, are available via a dedicated smartphone application (``Cosmic Ray Live'') developed for both Android and iOS systems (figure~\ref{fig:app}). This application enables the connection of several CRCs around the world, providing a global network for educational purposes. Presently, CRCs are located at: Gran Sasso National Laboratory (Italy), New York University of Abu Dhabi (UAE)~\cite{mannaei,arneodo}, INFN Sezione of Naples (Italy), and Canfranc Underground Laboratory (Spain).

\begin{figure*}[!th]
\centering
   \includegraphics[width=1.0\textwidth]{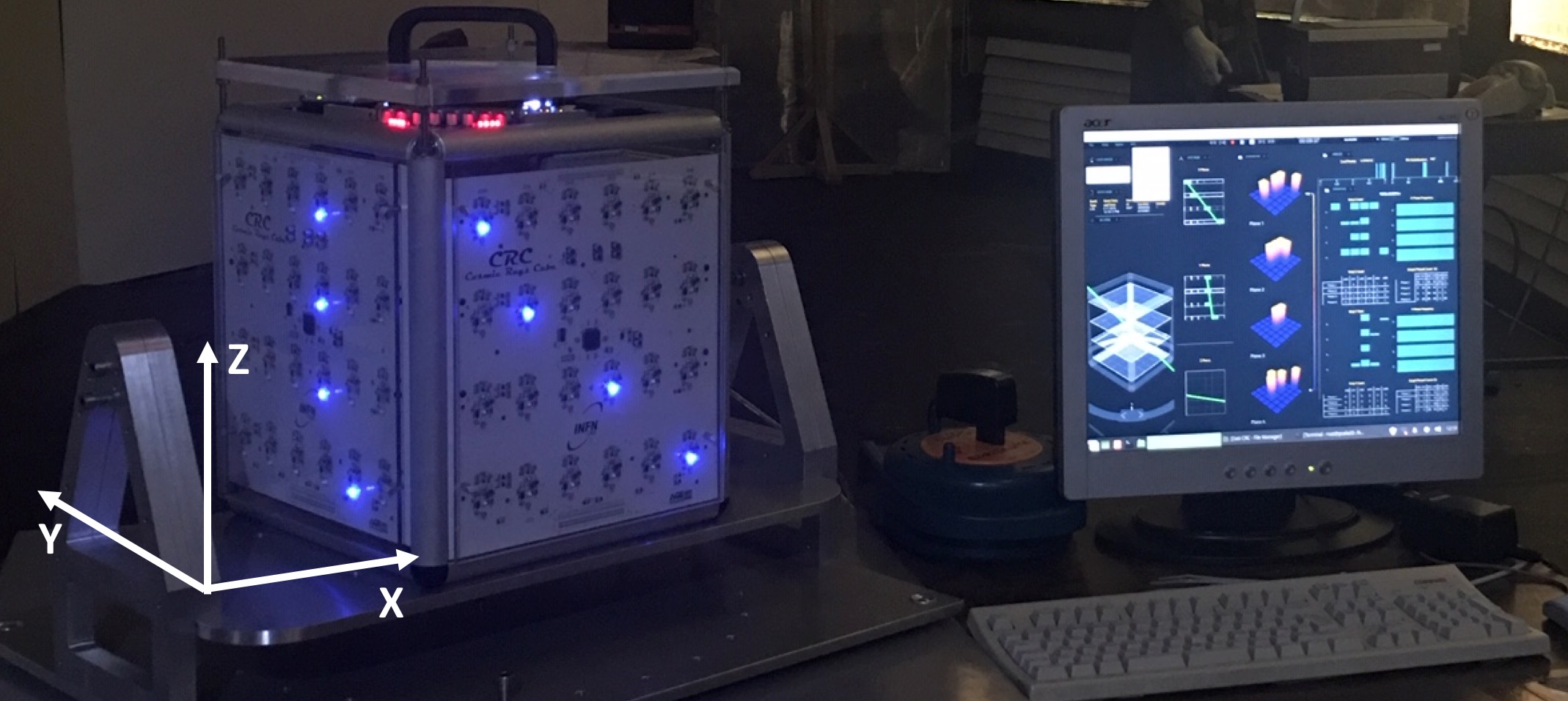}
   \caption{Photograph of the CRC detector operating in the ``war room'' of the underground ``Path of memory'' museum. A 3D track reconstruction of a muon crossing the hodoscope is well visible. The mechanical structure supporting the detector allows the rotation of the cube both in the polar and azimuth angles.}
   \label{fig:soratte}
\end{figure*}

\section{Event reconstruction and data analysis}
\label{sec:reco}

\begin{figure}[ht]
   \centering
   \includegraphics[width=1.0\columnwidth]{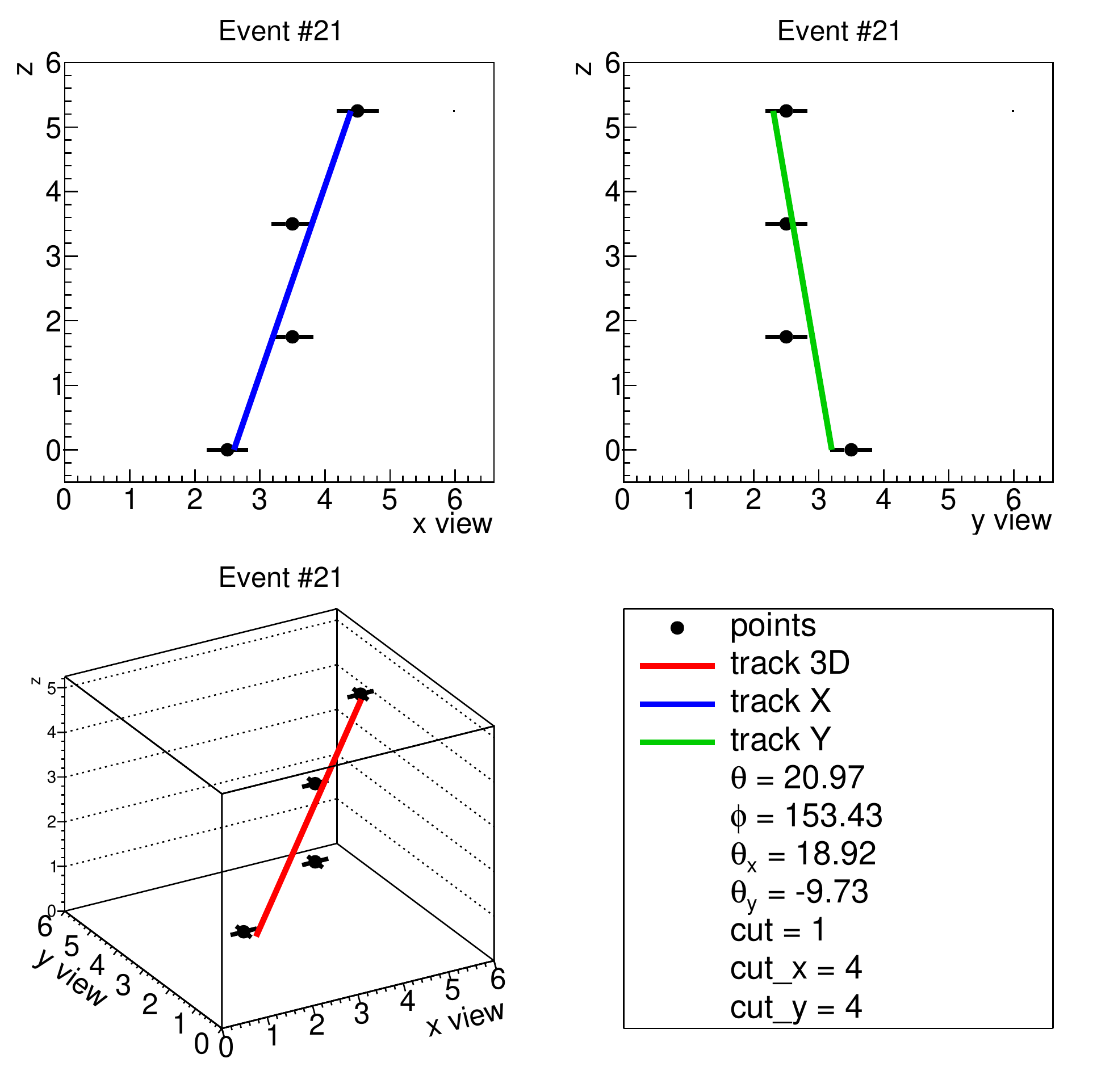}
   \caption{Example of event reconstruction. \emph{Top left and right}: reconstruction of the projections on the $x$ and $y$ views respectively. \emph{Bottom left}: 3D reconstruction of the muon track in the CRC  detector. \emph{Bottom right}: legend with angles of interest and information about the event selection.}
   \label{fig:ev}
\end{figure}

The 3-dimensional trajectories of muons crossing the detector are reconstructed from the 2-dimensional projections of the signals registered on the so-called $x$ and $y$ views, respectively, identified as the right and the left panels of the CRC detector (figure~\ref{fig:soratte}). The direction of each track is identified by the slope of the projected straight line in each view (figure~\ref{fig:ev}), determined with a linear regression algorithm based on the least squares method. 
Each point is plotted in the figure with its standard deviation set to $1/\sqrt{12}$ of the strip width (based on the uniform distribution hypothesis).

\begin{figure}[ht]
   \centering
   \includegraphics[width=1.0\columnwidth]{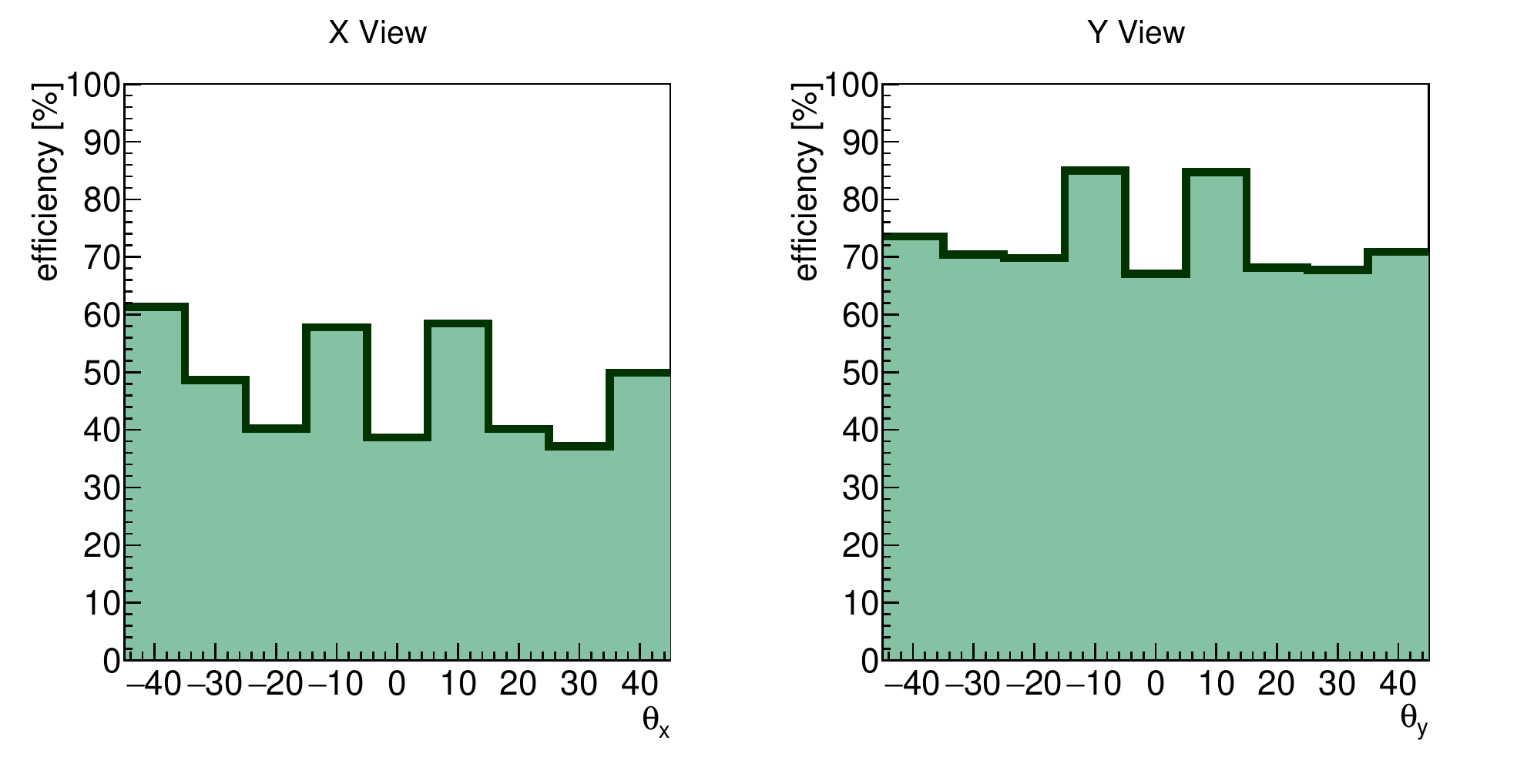}
   \caption{Efficiency for $x$ and $y$ views. The efficiency in percentage is reported as a function of the view angles $\theta_x$ and $\theta_y$ in 9 bins of 10$^\circ$ width in the interval (-45°, -45°). The shapes of the efficiency distributions reflect the preferred trajectories resulting from the segmented structure of the detector.
   }
   \label{fig:eff}
\end{figure}

\begin{figure*}[ht]
\centering
   \includegraphics[width=2.0\columnwidth]{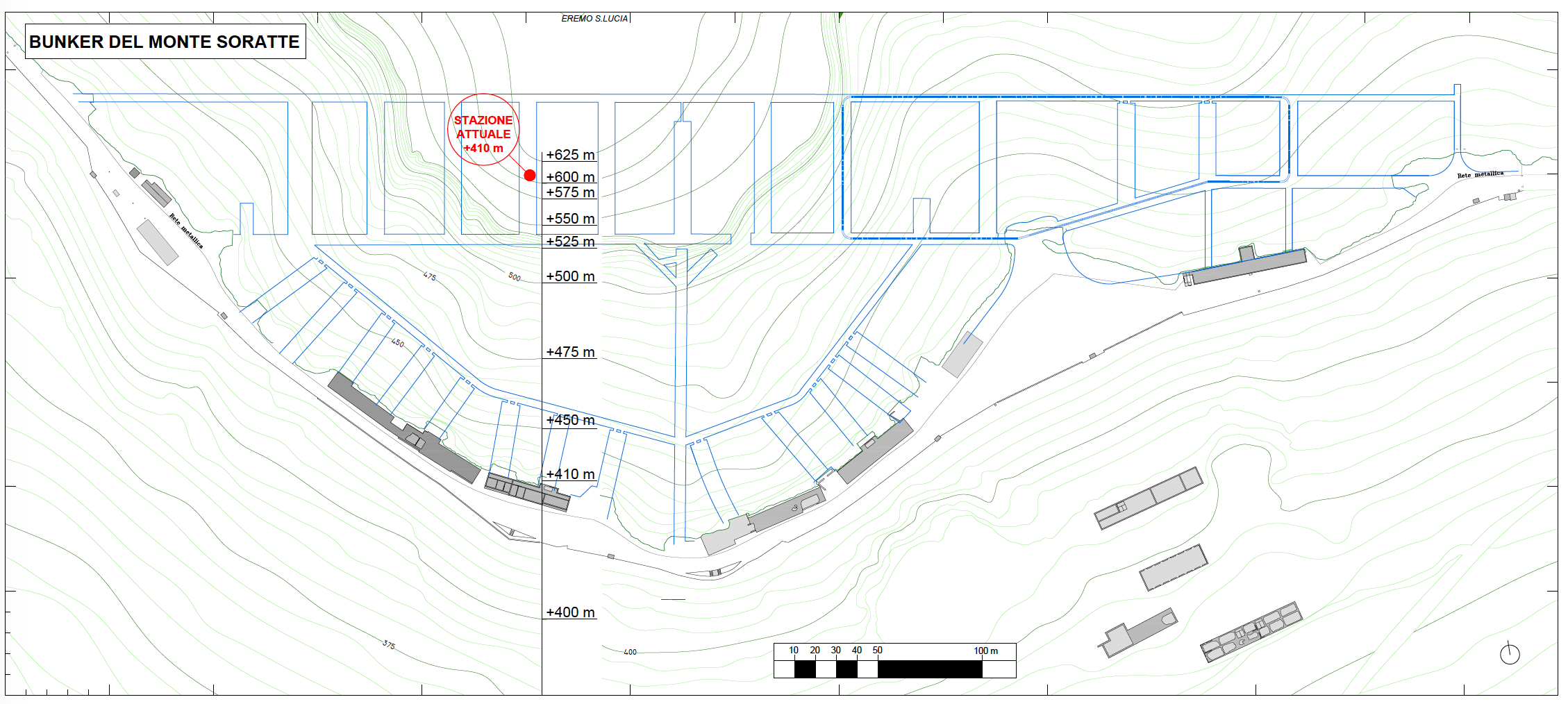}
   \caption{Planimetry of the Soratte bunker overlapped to the level contour lines of Mt. Soratte. The red dot indicates the position of the CRC detector inside the ``war room'' of the museum.}
   \label{fig:planim}
\end{figure*}

Good events are defined by the following selection criteria:
\begin{enumerate}
\item each of the four planes of a lateral view has at least one LED turned on;
\item if two LEDs are turned on in a plane, they must be contiguous;
\item the reduced $\chi^2$ of the linear regression must be lower than 1.5;
\item the conditions 1, 2 and 3 must be simultaneously valid for both the $x$ and $y$ views.
\end{enumerate}
The first two conditions ensure that the event can be safely associated with a clean muon, without sparse hits due to ambient radioactivity and/or electronic noise. Condition 3, suitably tuned with a Monte Carlo simulation, selects events produced by an ionising particle moving along a straight trajectory. Finally, the last condition requires that both CRCs views identify a clean track, allowing to reconstruct its 3-dimensional direction.

The two angles $\theta_x$ and $\theta_y$ are defined between the reconstructed track and the $z$ axis for each view (figure~\ref{fig:ev}).
The domain of both $\theta_x$ and $\theta_y$ is the fiducial angular interval $[-45^{\circ}, \, 45^\circ ]$. This range is close to the maximal angular acceptance determined by the cubic shape of the detector and by the selection criteria, requiring all the four planes of each view to be crossed by the particle. The entire angular range is divided into 9 bins of $10^\circ$ width.
Hence, tracks populate $9\times9$ bins corresponding to 81 different directions with respect to the $z$ axis of the CRC detector.
This provides a smooth view of the events angular distributions which match approximately the intrinsic angular resolution given by the strip width.

The geometrical acceptance, in terms of the $\theta_x$ and $\theta_y$ angles, is given by the product of the projected acceptances in the $x$ and $y$ views:
\begin{equation}
A(\theta_x, \theta_y) = A(\theta_x) A(\theta_y)
\end{equation}

\noindent The acceptance of both views $A(\theta_{x,y})$ only depends on the CRC geometry and on the adopted event selection, in which the four planes in each view are crossed by the particle. It can be expressed by the following formula:
\begin{equation}
    A(\theta_{x,y}) = \cos \theta_{x,y} \, \left(1 - \frac{Z_0}{L_0} \tan|\theta_{x,y}|\right) , 
\end{equation}

\noindent where $L_0 = 24$~cm and $Z_0 = 21$~cm are the width and height of the CRC, respectively (figure~\ref{fig:sketch}).

For the events fulfilling the selection criteria, the total efficiency is the product of the efficiencies of the 4 planes, both in the $x$ and $y$ views.
The CRC efficiency of each plane is measured on the surface as a function of the track slope, using $10^6$~events.

The efficiency of the $n_x$-th plane of the $x$ view with a given $\theta_x$ direction is defined as 
\begin{equation}
\epsilon(n_x,\theta_x) = N(n_x,\theta_x)/N_{tot}(n_x,\theta_x) ,
\end{equation}

\noindent where $N(n_x,\theta_x)$ is the number of events that fulfill the selection criteria for both views and belong to the $\theta_x$ angular bin, while $N_{tot}(n_x,\theta_x)$ also includes the events in which the $n_x$-th plane is inefficient (no LED on).

The typical efficiency of each of the $x$-view planes is around 84\%, leading to a 4-fold efficiency of about 50\% (figure~\ref{fig:eff}). Analogously, the efficiency of each of the $y$ planes is about 93\%, with a total view efficiency of 75\%.
The measured efficiencies, $\epsilon(\theta_x)$ and $\epsilon(\theta_y)$, are depicted in figure~\ref{fig:eff}.
The resulting total efficiency, also requiring the logical AND of the two views, $\epsilon(\theta_x,\theta_y)=\epsilon(\theta_x)\epsilon(\theta_y)$ is of the order of 35\%. 
This overall efficiency is the consequence of the high threshold setting and the AND logic adopted for the trigger, which significantly suppresses noisy events, avoiding ambiguities in the track reconstruction and providing a precise determination of the muon direction.

The adopted criteria allow us to measure the differential muon flux in a relatively short time with a statistical uncertainty adequate for our purposes.

The measured differential muon flux is defined as
\begin{equation}
    \Phi_{\mu}(\theta_{x},\theta_{y}) = \frac{n_{exp}}{\omega \cdot S \cdot A(\theta_{x,y})\epsilon(\theta_x,\theta_y)} 
\end{equation}
In this equation, $\Phi_{\mu}$ is the differential muon flux relative to the bin related to a specific direction defined by $(\theta_x, \theta_y)$. It is expressed in muons/s/sr. $n_{exp}$ is the muon rate relative to the angular bin, $\omega$ is the solid angle defined in a bin (equal to 0.0305 sr for our $\Delta\theta_x=\Delta\theta_y$ bin size), and $S=L_0^2=575~cm^2$ is the surface of a CRC plane. $A(\theta_{x,y})$ and $\epsilon(\theta_x,\theta_y)$ are the detector acceptance and efficiency, respectively.

\begin{figure}[ht]
   \centering
   \includegraphics[width=1.0\columnwidth]{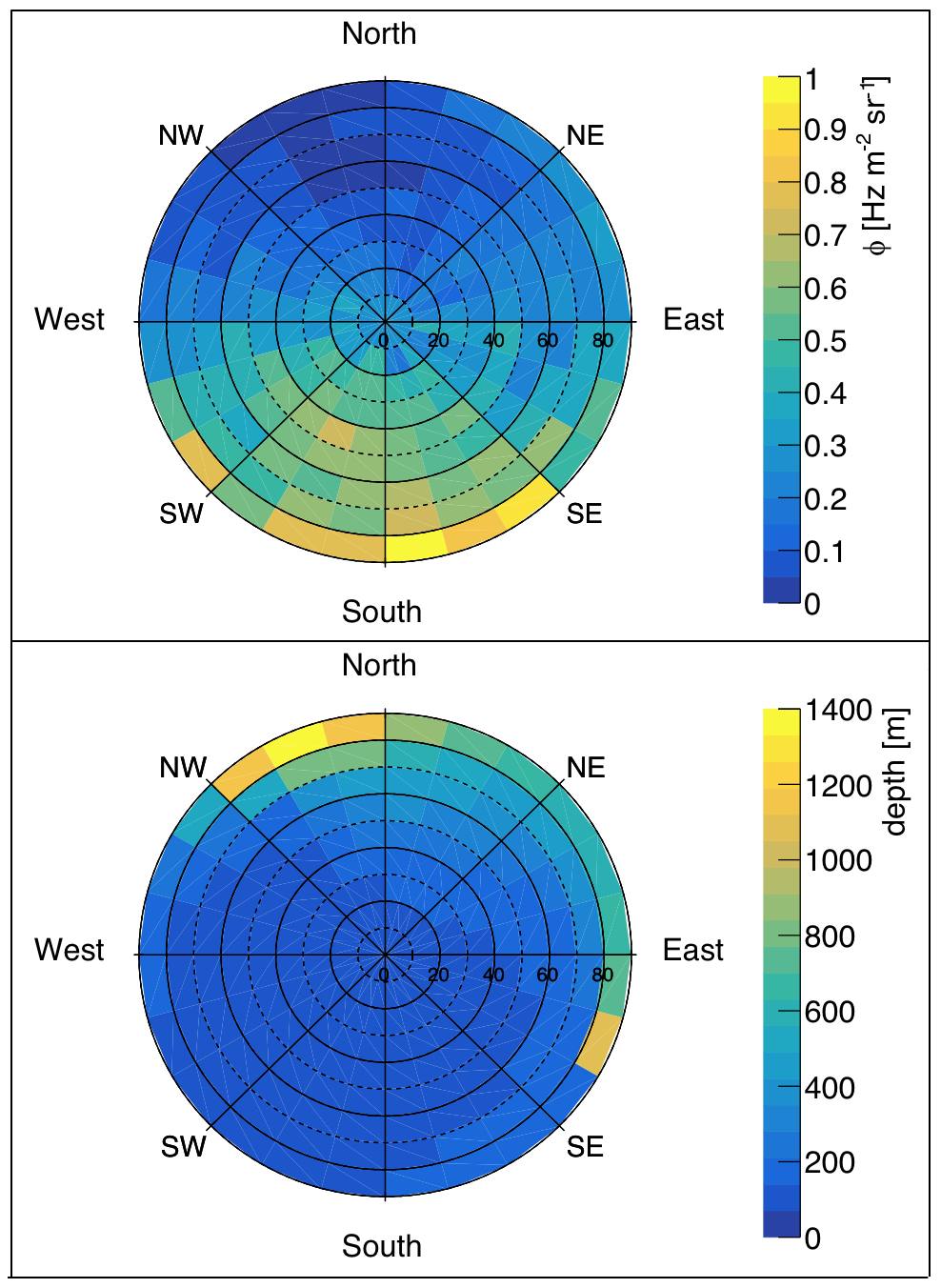}
   \caption{Top: Polar plot of the differential muon flux as measured in the ``war room'' of the Soratte bunker; bottom: slant depth of the Soratte mountain from the point of view of the CRC detector. The anticorrelation between the muon flux and the nominal thickness of rock crossed by the detected muons is well visible.}
   \label{fig:mappe}
\end{figure}

\begin{figure}[ht]
   \centering
   \includegraphics[width=1.0\columnwidth]{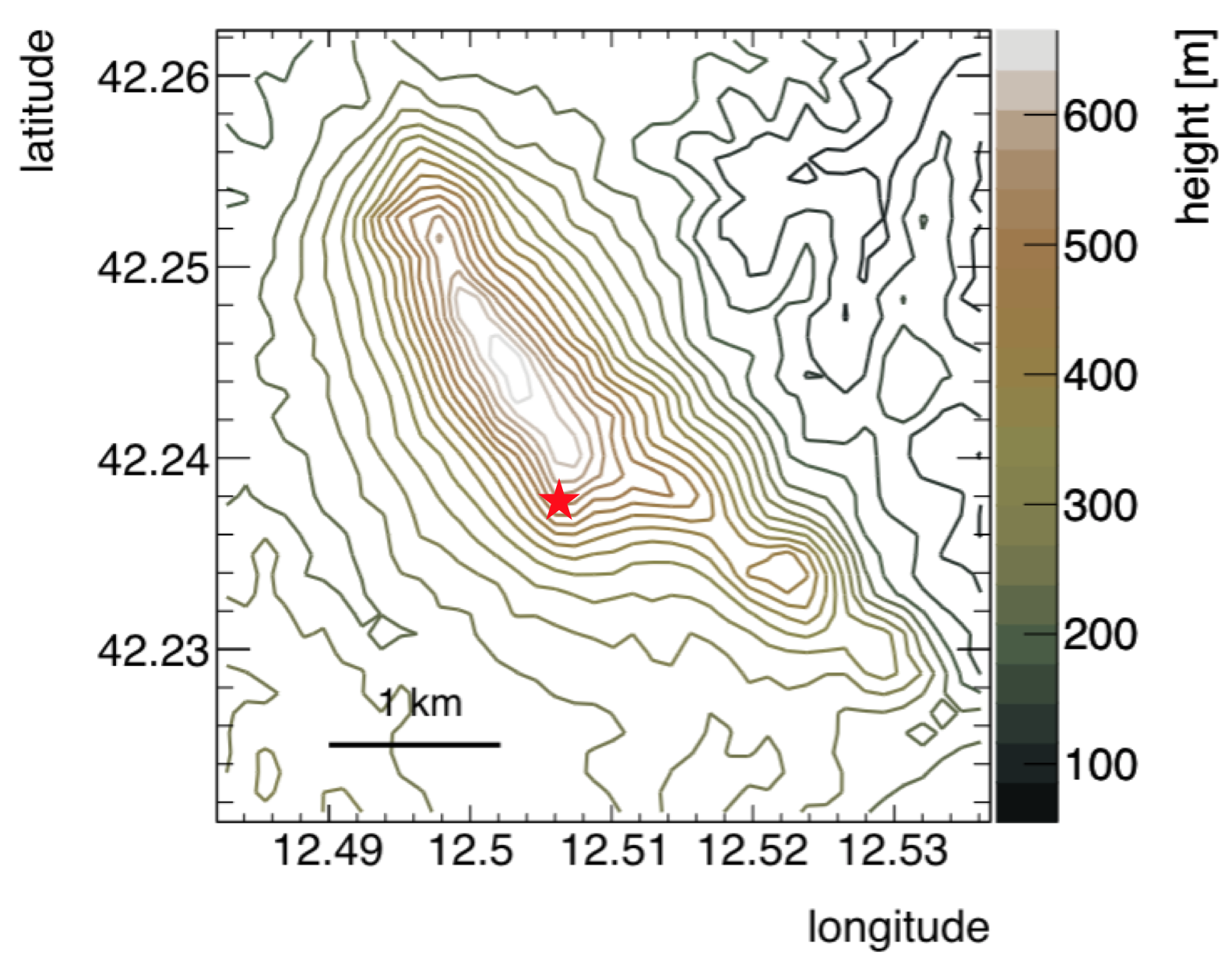}
   \caption{Topographic profile of Monte Soratte.}
   \label{fig:isosoratte}
\end{figure}

\section{Muon flux at Monte Soratte}
\label{sec:muon}

During the measurements, the detector was placed on a dedicated mechanical structure (figure~\ref{fig:soratte}) able to rotate the detector with respect to the z- and x-axes in angular steps of $\Delta\phi=15^{\circ}$ and $\Delta\theta=15^{\circ}$  ($\theta$ is the angle relative to the zenith, and $\phi$ is the azimuth angle). 
In this way, it was possible to overcome the limited angular acceptance of CRC (of about 45$^{\circ}$ with respect to its z-axis), making possible the measurement of the differential muon flux in the whole upper hemisphere.
To account for the CRC orientation, it is convenient to express the direction of the tracks in terms of standard polar angles, performing a change of variables and a proper trigonometrical rotation. In the following, terrestrial coordinates will be adopted, e.g. translating $\Phi_{\mu}(\theta_{x},\theta_{y})$ into $\Phi_{\mu}(\theta,\phi)$.
The underground measurements were performed using the following 43 orientations (angles in degrees):

\begin{enumerate}[(i)]
\item $\theta=0$, $\phi=0,45,180$;
\item $\theta=15$, $\phi=0,45,90,..,315$;
\item $\theta=30$, $\phi=0,45,90,..,315$;
\item $\theta=45$, $\phi=0,15,30,..,345$.
\end{enumerate}

Figure~\ref{fig:planim} shows the planimetry of the Soratte bunker and the position of the detector used to measure the differential muon flux. It is located close to the south side of the Soratte peak, at about 410 meters a.s.l. and covered by about 200~meters of rock along the vertical direction. 

The first polar plot in figure~\ref{fig:mappe} presents the differential muon flux of the upper hemisphere. The plot centre ($\theta=0^{\circ}$) indicates the vertical direction, while the external circle corresponds to the horizon ($\theta=90^{\circ}$). As expected, the differential muon flux is not homogeneous and reflects the side CRC positioning relative to the Soratte peak. The highest flux measured is close to the horizon and it comes from the southern sector, as expected. Instead, the horizontal muon flux corresponding to an azimuthal angle at about $\phi=338^{\circ}$ (clockwise relative to the north) is lower, as only muons with very high energy can cross more than 1~km of rock present along this direction. 
A low muon flux is measured also in the azimuth direction but at lower zenith angles. This behaviour can be easily understood by inspecting figure~\ref{fig:isosoratte}, which shows that the mountain ridge is just along this azimuthal direction. 
The overall anticorrelation of the differential muon flux with respect to the traversed rock thickness is also visible comparing the polar plot on the top with the second one, in which is reported the behaviour of the nominal rock thickness (the tunnels and halls of the Soratte bunker shown in figure~\ref{fig:planim} are neglected in this plot): as expected, the higher the rock thickness, the lower the muon flux. 
The total averaged flux is $\Phi_{\mu} = 0.3$~muons~s$^{-1}$st$^{-1}$m$^{-2}$, which is above 2 orders of magnitude lower than the flux measured on the surface, in agreement with our expectations.

\section*{Conclusions}

The reported measurement is the first of a campaign aimed at searching for a site with a good logistic and an affordable need of civil works to eventually host long-standing astroparticle physics experiments requiring low cosmic ray induced background.
Our result makes the Soratte bunker a suitable candidate to host \textsc{Ptolemy} and other astroparticle physics experiments that do not require a high muon flux reduction.

The CRC detector proved to be very suitable for the measurement discussed in this paper, as it can be remotely controlled, does not require special equipment and only needs little person-power.
These features were essential in the data taking performed during the COVID-19 pandemic lockdown period.
Given the present results, we are planning to add the Soratte bunker to the educational CRC network at disposal of high school students.

Thanks to the dedicated mechanical structure for vertical and horizontal rotation, it was possible to exploit the CRC features for a fine scanning of the differential muon rate, able to highlight the details of the mountain above the bunker. The total underground muon rate is found to be compatible with two orders of magnitude attenuation as expected from the average thickness of the rock. 

\section*{Acknowledgments}
We acknowledge the invaluable contribution of the mechanical workshop at LNGS. Also, we thank the staff of the ``Path of Memory'' museum for the warm hospitality and for helping with the installation and the operation of the detector. Finally, we thank Adriano Di Giovanni who helped with the proofreading.


\end{document}